\documentclass[conference]{IEEEtran}

\IEEEoverridecommandlockouts

\usepackage{cite}
\usepackage{amsmath,amssymb,amsfonts}
\usepackage{algorithmic}
\usepackage{graphicx}
\usepackage{textcomp}
\usepackage{xcolor}
\def\BibTeX{{\rm B\kern-.05em{\sc i\kern-.025em b}\kern-.08em
    T\kern-.1667em\lower.7ex\hbox{E}\kern-.125emX}}
\usepackage{booktabs}
\graphicspath{ {./plots/} }

\makeatletter
\newcommand{\linebreakand}{%
    \end{@IEEEauthorhalign}
    \hfill\mbox{}\par
    \mbox{}\hfill\begin{@IEEEauthorhalign}
}
\makeatother

\begin{document}

\begin{titlepage}
\thispagestyle{empty} 
\vspace*{\fill} 
\begin{center}
    \textbf{Copyright Statement}\\[1cm]
    This paper has been accepted for publication in the IEEE ITCA 2024 conference \\

    © 2024 IEEE. Personal use of this material is permitted. Permission from IEEE must be obtained for all other uses, in any current or future media, including reprinting/republishing this material for advertising or promotional purposes, creating new collective works, for resale or redistribution to servers or lists, or reuse of any copyrighted component of this work in other works.
\end{center}
\vspace*{\fill}
\end{titlepage}

\title{Cardiovascular Disease Detection By Leveraging Semi-Supervised Learning}

\author{
    \IEEEauthorblockN{Shaohan Chen\textsuperscript{1,*}, Zheyan Liu\textsuperscript{2a}, Huili Zheng\textsuperscript{2b}, and Qimin Zhang\textsuperscript{2c}, Yiru Gong\textsuperscript{2d}}
    \IEEEauthorblockA{
        \textsuperscript{1, 2}\textit{Department of Biostatistics, Columbia University, New York, NY 10032, USA}
    }
    \IEEEauthorblockA{
        shaohan.chen@caa.columbia.edu\textsuperscript{1,*}
        zl3119@caa.columbia.edu\textsuperscript{2a},\\
        hz2710@caa.columbia.edu\textsuperscript{2b,*}, 
        qimin.zhang@columbia.edu\textsuperscript{2c}
        yiru.g@columbia.edu\textsuperscript{2d},
    }
}

\maketitle

\begin{abstract}
Cardiovascular disease (CVD) persists as a primary cause of death on a global scale, which requires more effective and timely detection methods. Traditional supervised learning approaches for CVD detection rely heavily on large-labeled datasets, which are often difficult to obtain. This paper employs semi-supervised learning models to boost efficiency and accuracy of CVD detection when there are few labeled samples. By leveraging both labeled and vast amounts of unlabeled data, our approach demonstrates improvements in prediction performance, while reducing the dependency on labeled data. Experimental results in a publicly available dataset show that semi-supervised models outperform traditional supervised learning techniques, providing an intriguing approach for the initial identification of cardiovascular disease within clinical environments.
\end{abstract}


\section{Introduction}
Cardiovascular disease (CVD) ranks among the most common and lethal health issues globally, accounting for a significant proportion of mortality and morbidity each year. Swift detection and management of cardiovascular disorders are pivotal for enhancing patient results and reducing pressures on healthcare frameworks.

Supervised learning techniques have been widely applied in the healthcare fields\cite{zheng2024identificationprognosticbiomarkersstage}, showing promise in diagnosing and predicting various health conditions\cite{zhang2024comparativeanalysisofbeyesian}, including CVD. However, The efficacy of early detection techniques is frequently constrained by the availability of high-quality labeled data, which is essential for training supervised machine learning models\cite{gu2023identification}. The scarcity of labeled data is a particularly challenging issue in the healthcare domain\cite{zhai2024long}, where expert annotations are often required. Consequently, the performance of supervised models may be hindered by the limited availability of labeled data, potentially leading to sub-optimal diagnostic accuracy\cite{kang2022tie}.

In contrast, semi-supervised learning offers a compelling alternative by utilizing both labeled and unlabeled data, making it possible to improve model performance even with a relatively small amount of labeled data\cite{hu2023artificial}. Semi-supervised learning approaches have been effectively implemented in various domains\cite{2411.06067}, yet their potential in the realm of cardiovascular disease detection remains under-explored.

In this paper, we enhance CVD detection using semi-supervised learning. The approach we propose integrates labeled patient data with a significant amount of data without label, employing semi-supervised learning techniques to enhance the dependability of cardiovascular disease detection.We perform experiments on publicly accessible datasets, showcasing the efficacy of semi-supervised learning approach compared to supervised learning methods.

The subsequent portions of our document are structured thus: Section II delineates the use of datasets and our methods for preprocessing. Section III explicates the techniques utilized. Section IV reports on the experimental framework and findings. Section V offers conclusions and implications, and Section VI explores potential directions for future research.

\section{Data}
The dataset utilized in our experiments is from the Behavioral Risk Factor Surveillance System (BRFSS), a major ongoing healthcare survey \cite{pierannunzi2013systematic} which gathers data on healthcare stratified risk behavior and behavior changes, chronic health condition with maintenance medication, and utilization rate of proactive services, primarily via telephone interviews.

\subsection{Variables}
In out dataset, there are 19 variables been extracted that relates to lifestyle factors of a person that can be contributed to being at risk with any form of CVD. The response variable is \texttt{Heart Disease}, shows whether a person has been diagnosed with heart disease, denoted by "Yes" or "No". Predictors include other health factors of the individual including: \texttt{routine checkup frequency}, \texttt{height}/\texttt{weight}/\texttt{BMI}, \texttt{alcohol/fruit consumption}, \texttt{depression history}, \texttt{diabetes history}, etc.

\subsection{Preprocessing}

Samples with any missing value are checked and removed from the dataset. The response variable is converted from categorical ("Yes", "No") to binary (1, 0). All categorical predictors were identified and transformed into dummy variables using one-hot coding. This process converts each category into separate binary columns, ensuring the data is numeric and suitable for model training.

After processing, the dataset contains only numeric variables, including the binary response and encoded features\cite{jin2024learning}. The subset sample size after sampling from original dataset is $20,544$. This cleaned dataset is now ready for model analysis.





\subsection{Data Workflow}

\begin{figure}[htbp]
\centerline{\includegraphics[width=0.3\textwidth]{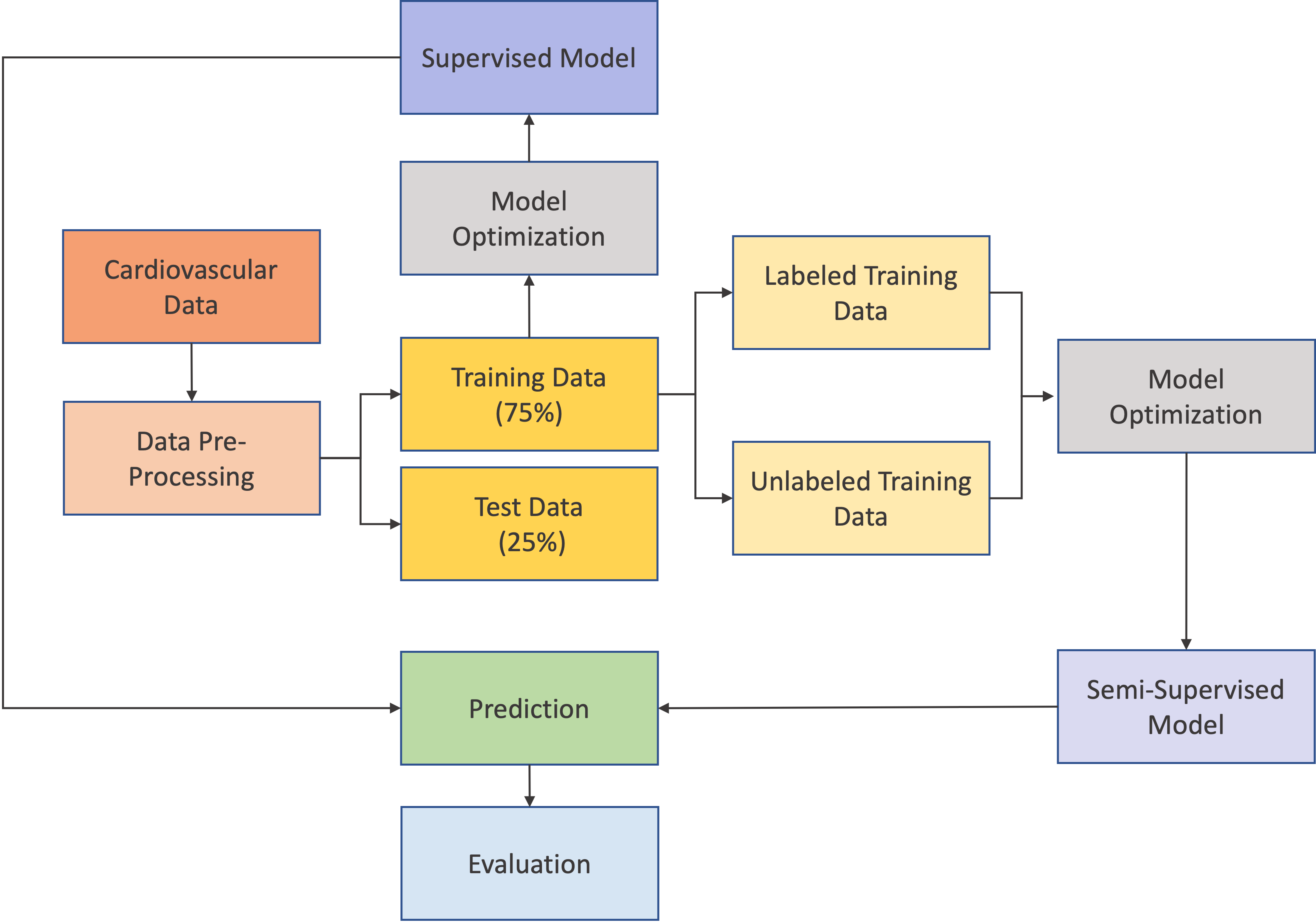}}
\caption{Data Workflow}
\label{data_flow}
\end{figure}

The overview of data work flow is illustrated in the flowchart (Figure \ref{data_flow}).
Initially, we preprocess the cardiovascular dataset to ensure data cleanliness and split it into 75\% for training and 25\% for testing. A copy of the original training data is retained for supervised learning. The training data is subsequently split into labeled and unlabeled subsets. For supervised learning models, the entire training set is utilized for model training and parameter optimization, while a mix of both labeled and unlabeled data will be utilized for semi-supervised learning models. Ultimately, we assess and contrast the performance of both types of models using the test dataset.

\section{Methodology}

\subsection{Semi-Supervised Learning Models}
Our study utilizes five semi-supervised learning approaches to effectively harness both labeled and unlabeled data:

\begin{itemize}
    \item \textbf{Semi-Supervised SVM:}
    This method extends the Support Vector Machine (SVM) framework by maximizing the margin between classes using labeled data, while also taking into account the distribution of unlabeled data\cite{zhu2023demonstration}. The objective function can be represented as:
    \[
    \min_{\mathbf{w}, b, \xi} \frac{1}{2} \|\mathbf{w}\|^2 + C \sum_{i=1}^l \xi_i + C^* \sum_{j=1}^u \xi_j
    \]
    where $\mathbf{w}$ as weight and $b$ as bias, $\xi_i$ and $\xi_j$ are the slack variables for labeled and unlabeled data respectively, $C$ and $C^*$ are penalty parameters.

    \item \textbf{Self-Training:}
    This model begins with a supervised model (here we employ logistic regression, which is subsequently utilized to predict labels for the unlabeled data. Only the most confident predictions (based on a confidence threshold) are selected, and the corresponding pseudo-labeled data points are added to the training set\cite{huang2024ar}. The process is repeated iteratively:
    \[
    \hat{y} = \text{argmax}_y P_\theta(y | x), \quad \text{if } P_\theta(y | x) > \text{threshold}
    \]
    where $P_\theta(y|x)$ is the prediction probability of label $y$ given input $x$.

    \item \textbf{Pseudo-Labeling:}
    Pseudo-Labeling is a variant of Self-Training that specifically focuses on adding the most confident predictions from the unlabeled data to the training set. Unlike Self-Training, Pseudo-Labeling typically involves directly setting a high-confidence threshold\cite{kang20216}:

    \item \textbf{Mean Teacher:}
    Mean Teacher is an advanced semi-supervised learning model that operates with two neural networks: a network called 'student' and a teacher named as 'teacher', where the weight of the latter network are derived as an Exponential Moving Average (EMA) of the former network's weights.\cite{Li2024.09.08.24313212}:
    \[
    \theta_{\text{teacher}}^{(t)} = \alpha \theta_{\text{teacher}}^{(t-1)} + (1 - \alpha) \theta_{\text{student}}^{(t)}
    \]
    where $\theta_{\text{student}}$ and $\theta_{\text{teacher}}$ represent distinctively the weight of the student and teacher network.

    \item \textbf{$\Pi$-Model:}
    $\Pi$-Model is commonly applied in neural networks\cite{zhu2022optimizing}, is designed to utilize both labeled and unlabeled data by enforcing consistency regularization:
    \[
    \mathcal{L} = \mathcal{L}_{\text{sup}} + \lambda \mathcal{L}_{\text{cons}}(f(x), f(x'))
    \]
    where $x'$ is a perturbed version of $x$, $f$ represents the neural network function, $\mathcal{L}_{\text{sup}}$ is the supervised loss, and $\mathcal{L}_{\text{cons}}$ is the consistency loss.
\end{itemize}

\subsection{Supervised Learning Models}
For comparison, we also adopt five different supervised learning models:
\begin{itemize}
    \item \textbf{Logistic Regression:} Logistic Regression is a linear method designed for and applied in classification scenarios, calculating the probability \( p \) that the input \( \mathbf{x} \) belongs to a class labeled as 1. The model uses the logistic function\cite{gu2024identification}:
    \[
    p = \frac{1}{1 + e^{-\mathbf{w}^T \mathbf{x}}}
    \]

    \item \textbf{Decision Tree:} This is a tree-based model that splits data based on features to create branches, leading to leaf nodes that represent class labels\cite{zhu2021twitter}. Each decision node in the tree represents a feature \( x_i \) and a threshold \( t \), splitting the data into two subsets:
    \[
    \text{if } x_i \leq t \text{ then left child, else right child}
    \]

    \item \textbf{Random Forest:} Random Forest, a collective learning strategy, constructs multiple decision trees using varied subsets of data and attributes.\cite{yan2022influencing}. The ultimate classification decision emerges from a predominant consensus among these trees.
    
    \item \textbf{K-Nearest Neighbors (KNN):} allocates a label to a data point according to the most common label among its closest neighbors\cite{202410.1536}. The parameter \(k\), which denotes the number of neighbors, affects the smoothness of the decision boundary.

    \item \textbf{Gradient Boosting:} Gradient Boosting is one of the ensemble techniques that constructs numerous decision trees in a sequence, in which each subsequent tree aimed at correcting the errors made by the previous decision tree\cite{su2022mixed}. If \( F_m \) is the model built after \( m \) trees, the update rule is:
    \[
    F_{m+1}(\mathbf{x}) = F_m(\mathbf{x}) + \gamma h_m(\mathbf{x})
    \]
    where \( h_m \) is the new decision tree and \( \gamma \) is the learning rate.
\end{itemize}

\section{Experimental Results}

In this part of the paper, we will detail the results of our experiments with semi-supervised learning. Our focus was on evaluating various models in terms of AUC and F1-score for different proportions of labeled data, and comparing them against supervised learning methods\cite{202411diabetesprediction}.

\subsection{Semi-Supervised Models' Performance with Labeled Data Proportions}

\begin{figure}[htbp]
\centerline{\includegraphics[width=0.3\textwidth]{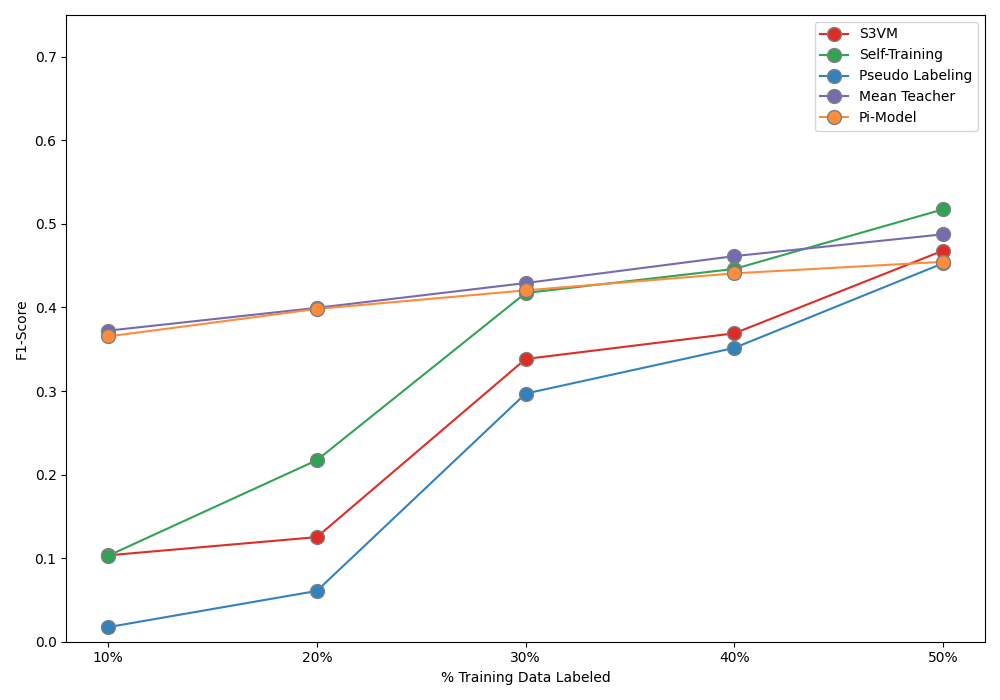}}
\caption{F1-Score of Semi-Supervised Models }
\label{semi-f1}
\end{figure}

\begin{figure}[htbp]
\centerline{\includegraphics[width=0.3\textwidth]{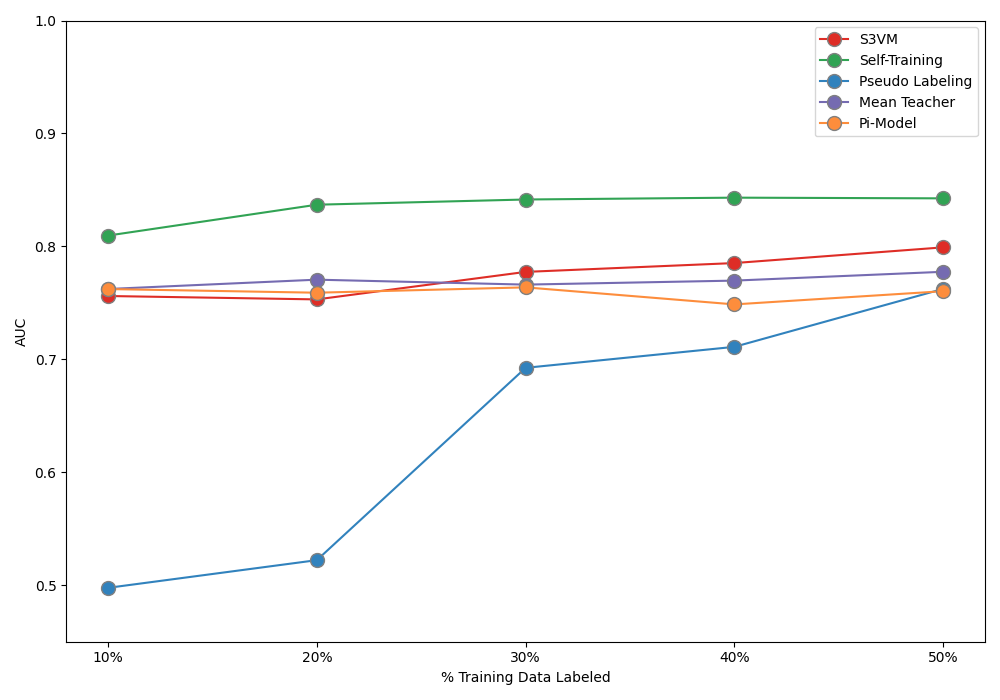}}
\caption{AUC of Semi-Supervised Models}
\label{semi-auc}
\end{figure}

\begin{table}[h!]
\centering
\caption{Semi-Supervised versus Supervised Models}
\label{tab:performance_comparison}  
\begin{tabular}{l c c c}
\toprule
\textbf{Model} & \textbf{Accuracy} & \textbf{F1-Score} & \textbf{AUC} \\
\midrule
\multicolumn{4}{l}{\textbf{\textit{Semi-Supervised Models (50\% training data labeled)}}} \\  
Semi-Supervised SVM    & 0.7930 & 0.4681 & 0.7992 \\
\textbf{Self-Training}          & \textbf{0.8003} & \textbf{0.5175} & \textbf{0.8425} \\
Pseudo-Labeling        & 0.7904 & 0.4530 & 0.7624 \\
Mean Teacher           & 0.7591 & 0.4876 & 0.7775 \\
Pi-Model               & 0.7375 & 0.4547 & 0.7603 \\
\midrule
\multicolumn{4}{l}{\textbf{\textit{Supervised Models}}} \\  
Logistic Regression    & 0.7986 & 0.5138 & 0.8443 \\
Random Forest          & 0.7898 & 0.4729 & 0.8193 \\
Decision Tree          & 0.7192 & 0.4554 & 0.6382 \\
KNN                    & 0.7110 & 0.1820 & 0.5521 \\
\textbf{Gradient Boosting}      & \textbf{0.8027} & \textbf{0.5179} & \textbf{0.8439} \\
\bottomrule
\end{tabular}
\end{table}

Figures \ref{semi-f1} and \ref{semi-auc} show performance enhancements in F1-Score and AUC metrics with increasing labeled data from 10\% to 50\%. Significant improvements in F1-Score across models highlight their improved precision and recall, especially as the proportion of labeled data grows. Initially low metrics in precision and recall saw notable advances, demonstrating the models' effective exploitation of both labeled and unlabeled data.

AUC values, crucial for assessing model reliability and generalization\cite{yang2024analysisfinancialriskbehavior}, remained consistently above $0.75$ for most models with over 30\% labeled data and often exceeded $0.8$. This trend, particularly strong in Self-Training, signifies robust classification performance across varying data availabilities. Overall, the data points to semi-supervised learning as an effective strategy in scenarios with limited access to fully labeled datasets, maintaining strong performance despite data constraints.

\subsection{Semi-Supervised Learning vs. Supervised Learning}

The comparison between semi-supervised and supervised learning models, as outlined in Table \ref{tab:performance_comparison}, reveals distinct performance levels. Among the semi-supervised models, Self-Training excels, achieving the highest Accuracy ($0.8003$), F1-Score ($0.5175$), and AUC ($0.8425$) with only 50

In supervised learning, Gradient Boosting led with the highest Accuracy ($0.8027$) and AUC ($0.8439$), closely rivaling the F1-Score of Self-Training. Logistic Regression displayed strong AUC performance ($0.8443$), while simpler models such as the Decision Tree and KNN lagged significantly behind in key metrics.

This analysis underscores that semi-supervised techniques, especially Self-Training, can match the top supervised models' metrics with half the labeled data, demonstrating their potential to reduce labeling requirements while maintaining accuracy, particularly beneficial in fields like cardiovascular disease detection.

\section{Conclusion}

In conclusion, the study demonstrates the effectiveness of semi-supervised learning models, particularly Semi-Supervised SVM and Self-Training, in low-labeled data environments. These models perform comparably, and in some cases, better than traditional supervised methods, especially when labeled data is scarce\cite{zhang2024optimizationapplicationcloudbaseddeep}. Semi-supervised learning proves to be a promising approach for tasks where labeling data is challenging or expensive\cite{zhou2024adapi}. Both Semi-Supervised SVM and Self-Training models achieve competitive AUC and F1-scores with 50\% labeled data, showcasing their strength in leveraging unlabeled data to improve classification performance.

\section{Discussion}

Future research can explore several avenues to further these findings. One promising direction is investigating the performance of other advanced semi-supervised methods, such as the Mean Teacher Model and Pi-Model, which focus on consistency regularization\cite{zhang2024cunetunetarchitectureefficient}. These models can further enhance generalization by enforcing consistency between predictions on perturbed versions of the input\cite{penglingcn}
.

In addition, incorporating unsupervised feature learning techniques, such as auto-encoders\cite{wu2024research} or contrastive learning\cite{zhong2024deep}, could complement existing semi-supervised approaches. By learning more robust representations from unlabeled data, these methods can reduce the reliance on labeled data and enhance performance further\cite{Shen2024Harnessing}.

Another promising research direction is integrating graph-based models, such as Label Propagation and Label Spreading, with current techniques\cite{xie2023accel}. These methods exploit the inherent structure in the data and, when combined with semi-supervised models, may yield even better performance\cite{10735529}.

Lastly, hybrid approaches that combine semi-supervised learning with transfer learning could be explored. This combination is especially relevant in domains where labeled data is scarce and domain shift issues are prevalent\cite{peng2024maxk}. These hybrid models may generalize better across domains with minimal labeled data.






\bibliographystyle{plain}
\bibliography{output.bib}

\end{document}